\documentclass[prl,aps,amsfonts,twocolumn,dvips]{revtex4}
\usepackage{graphics,epsfig,amssymb}

\usepackage{graphicx}
\usepackage{dcolumn}
\usepackage{bm}
\usepackage{amsmath}

\begin{document}
\title
{\large\bf Suppression of Superfluidity upon Overflow of Trapped Fermions. 
           Quantal and Thomas-Fermi Studies}

\author{P. Schuck$^{1,2}$, X. Vi\~nas$^{3}$}

\bigskip

\address{\rm
$^{1}$Institut de Physique Nucl\'eaire, IN2P3-CNRS, Universit\'e Paris-Sud,
\\F-91406 Orsay-C\'edex, France \\
$^{2}$ Laboratoire de Physique et Mod\'elisation des Milieux Condens\'es,
CNRS and Universit\'e Joseph Fourier, 25 Avenue des Martyrs, Bo\^{i}te Postale 166,
F-38042 Grenoble Cedex 9, France\\
$^{3}$Departament d'Estructura i Constituents de la Mat\`eria
and Institut de Ci\`encies del Cosmos, Facultat de F\'{\i}sica,
Universitat de Barcelona, Diagonal {\sl 647}, {\sl E-08028} Barcelona,
Spain \\
}

\begin{abstract}
Two issues are treated in this work: (i) the generic fact that if a 
fermionic superfluid in the BCS regime overflows from a narrow container 
into a much wider one, 
pairing is much suppressed at the overflow point. Physical examples where this 
feature may play an important role are discussed. (ii) A Thomas-Fermi (TF)
approach to inhomogeneous superfluid Fermi-systems is presented and shown that 
it works well in cases where the Local Density Approximation (LDA) breaks 
down.
\end{abstract}

\maketitle



Superfluid fermions in finite systems can exist in traps of cold 
atoms, in nuclear systems, in small metallic clusters, etc. An interesting 
question arises what 
happens to the superfluid if its Fermi level reaches the edge of a finite 
container, 
i.e. either the fluid overflows into the continuum or it pours into another 
container 
of much larger dimension. A trapping potential of this type has experimentally 
already been generated for the study of cold bosonic atoms \cite{kett98}. 
It should also be possible to use it for fermionic atoms \cite{viv01}.

In the inner crust of neutron stars, 
there also may occur the situation where coexists a superfluid 
neutron gas of variable density in between the lattice of (superfluid) 
nuclei \cite{bal07,cha10}. This situation often is mimicked by a Wigner Seitz 
cell, where the single particle potential has a pocket, representing the 
nucleus, embedded in a large container.
Still other systems may exist with similar situations.
\begin{figure}
\includegraphics[height=6.5cm,angle=-90]{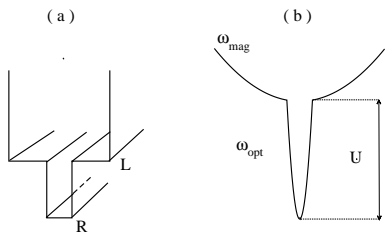}
\caption{
\label{Figure1} Schematic view of the potentials used in this work. Panel (a) 
shows a prespective view of the slab potential which is 
translationally invariant in x, y direction. Panel (b) represents a 
spherically symmetric container}
\end{figure}

The purpose of this work is to study superfluidity of fermions at the 
overflow (drip) in the BCS regime. 
Since the quantal solution of BCS equations in geometries with rapidly varying 
single particle potentials with a large number of particles is numerically 
difficult, 
we will present, as a second objective of this work, a Thomas-Fermi 
approach to inhomogeneous superfluidity 
which shows good performance 
in situations where LDA  fails. 

For our study we first will use a schematic 
model of slab geometry with a transverse potential of large extension $L$ 
posessing 
at the origin a 'pocket' of variable depth and size $R$ much smaller than the 
outer container. Schematically such a potential is shown in Fig.~1a. This 
slab configuration may roughly mock up one sheet of a 
so-called Lasagne 
configuration in the inner crust of neutron stars \cite{hae07}. We, 
therefore, will 
use nuclear dimensions for the slab model but they can easily be replaced by 
dimensions relevant 
for other systems. Our model and the ensuing generalisations treated below, 
therefore, are believed to be generic. We will 
study the slab 
configuration also because the quantal solution of the gap equation is 
evaluated relatively directly and the quality of the TF approach can thus 
be established. Once this is achieved, we also will go over to other 
geometrical configurations. We, for instance, will treat a second potential 
shown in Fig.~1b with spherical symmetry, a kind of which, as already 
mentioned, has been used for bosonic atoms in \cite{kett98}.

The wave 
functions and eigenenergies of a box as shown in Fig.~1a with a 
potential-hole are given 
in \cite{flu74}. For pairing, we use a contact force with a cut off $\Lambda$, 
to make things simple.
Integrating over momenta in slab direction the usual gap equation 
$\Delta_n= \sum_{n'} \int \frac{d^2p}{(2\pi \hbar)^2} V_{nn'}\Theta(\Lambda - \varepsilon_{n'} - \varepsilon_p) \Delta_{n'}/(2E_{n'}(p))$ with $E_n(p) = \sqrt{(\varepsilon_n + \varepsilon_p - \mu)^2 + \Delta_n^2}$ the quasiparticle energy, $\Theta(x)$ the step function, and $\varepsilon_n, \varepsilon_p$ being the discrete 
single particle energies in transverse direction and kinetic energies in 
slab direction, respectively,
one arrives at the following gap equation
 

\begin{equation}
\Delta_n = - \sum_{n'} \Theta(\Lambda - \varepsilon_{n'}) V_{nn'} K_{n'}
\label{eq1} \end{equation}
with

\begin{equation}
K_n =\frac{m}{4\pi\hbar^2}\Delta_{n} \ln{\frac{\Lambda - \mu + 
\sqrt{(\Lambda - \mu)^2 + 
\Delta_n^2}}
{\varepsilon_n - \mu + \sqrt{(\varepsilon_n - \mu)^2 + \Delta_n^2}}}
\label{eq2} \end{equation}

\begin{figure}
\includegraphics[height=7.5cm,angle=-90]{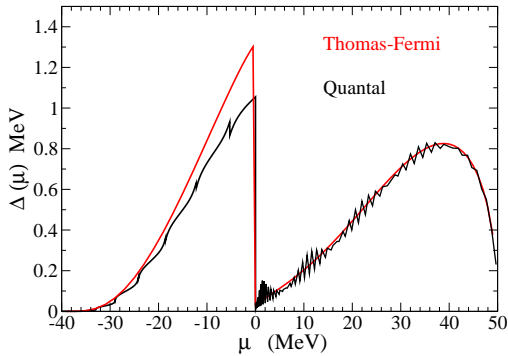}
\caption{(Coloronline) \label{Figure2} Quantal and TF pairing gap in 
the slab geometry as a
function of the chemical potential.}
\end{figure}

\noindent
where $m$ is the particle mass and  the indices $n$ stand for the level 
quantum numbers in the confining 
potential of Fig.~1a. 
The matrix elements 
$V_{nn'}= -g \int_{-L}^{+L}|\varphi_n(z)|^2 |\varphi_{n'}(z)|^2 dz$ of the 
contact force 
$-g\delta({\bf r} - {\bf r}')$ can be evaluated straightforwardly from 
the wave functions $\varphi_n(z)$ given in \cite{flu74}.

For an example we take as cut off $\Lambda$ = 50 MeV counted from the edge 
of the pocket potential whose depth be $V_0=-$40 MeV.
Its extension ranges from  $-R$ to +$R$ with $R$ = 10 fm. The wide potential 
with infinite walls has extension from $-L$ to +$L$ with $L$ = 100 fm. 
For the coupling strength we take $g$= 150 MeV fm$^3$.

Before we show the results, let us explain our Thomas-Fermi (TF) 
approach for this problem. 
For this, 
we transform Eq.(1) into a continuum version in the following way. We first 
consider the Wigner transform of the density matrix corresponding to the state 
$|n\rangle $: $[\hat \rho_n]_W = [|n\rangle \langle n|]_W$ and take 
the $\hbar \rightarrow 0$ limit of this expression, see \cite{vin11,vin03,RS}

\begin{equation}
f_E(z,p)=\frac{1}{g^{TF}(E)}\delta(E-H_{cl.}) + O(\hbar^2)
\label{eq3} \end{equation}

\noindent
where $H_{cl.} = \frac{ p^2}{2m} + V(z)$ is the classical Hamiltonian 
with $V(z)$ the potential (Fig.~1a) in transverse direction 
and $g^{TF}(E)$ is the corresponding level density to lowest order 
in $\hbar$, i.e. the usual
TF expression \cite{RS}. Quantum numbers and energies are simply characterised 
by the continuous energy variable $E$  which takes the place of the discrete 
values $\varepsilon_n$ in the quantal case.


The TF version of the gap equation (1) then reads

\begin{equation}
\Delta(E) = - \int_{V_0}^{\Lambda} dE' g(E') V(E,E') K(E')
\label{eq4} \end{equation}
with $K(E)$ an obvious generalisation of $K_{n}$.
The matrix elements $V(E,E')$ can be evaluated in replacing $|\varphi_n(z)|^2$
by \cite{vin03}
$
\rho^{TF}_E(z) = \frac{1}{g^{TF}(E)}\frac{1}{2\pi}
\big(\frac{2m}{\hbar^2} \big)^{1/2}
[E - V(z)]^{-1/2}$,
the on-shell TF density in transverse direction. As the reader will easily 
realise, 
the way of proceeding is very different from usual LDA where the finite size 
dependence is put into the (local) chemical potential whereas here it is 
put into the matrix elements of the pairing force.

\begin{figure}
\includegraphics[height=7.5cm,angle=-90]{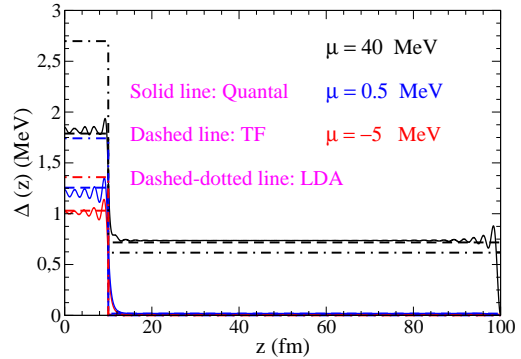}
\caption{(Coloronline) \label{Figure3} Position dependence of the gap in 
the slab geometry
for different values of the chemical potential. Quantal, TF, and LDA results 
are shown. Notice that $\Delta$ for 
$\mu$ = 0.5 and -5.0 MeV is practically zero in the gas region.}  
\end{figure}


\begin{figure}
\includegraphics[height=7.5cm,angle=-90]{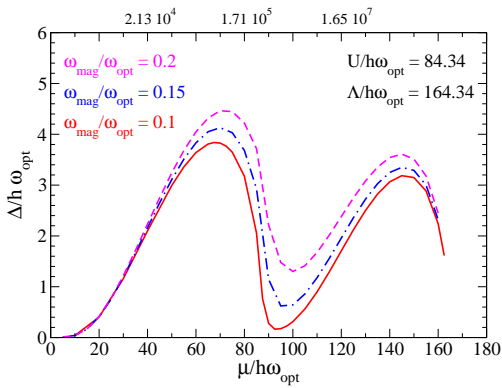}
\caption{(Coloronline) \label{Figure4} Average TF gaps at the Fermi energy 
as a function of 
the chemical potential for the potential shown in Fig. 1b. In the completely filled
optical trap ($\mu=U$) we accomodate 10$^5$ atoms in each spin state.
The total number of atoms in the trap with
$\mu/\hbar\omega_{opt}$=40, 80 and 120 are indicated in the upper 
horizontal axis.} 
\end{figure}



\begin{figure}
\includegraphics[height=7.5cm,angle=-90]{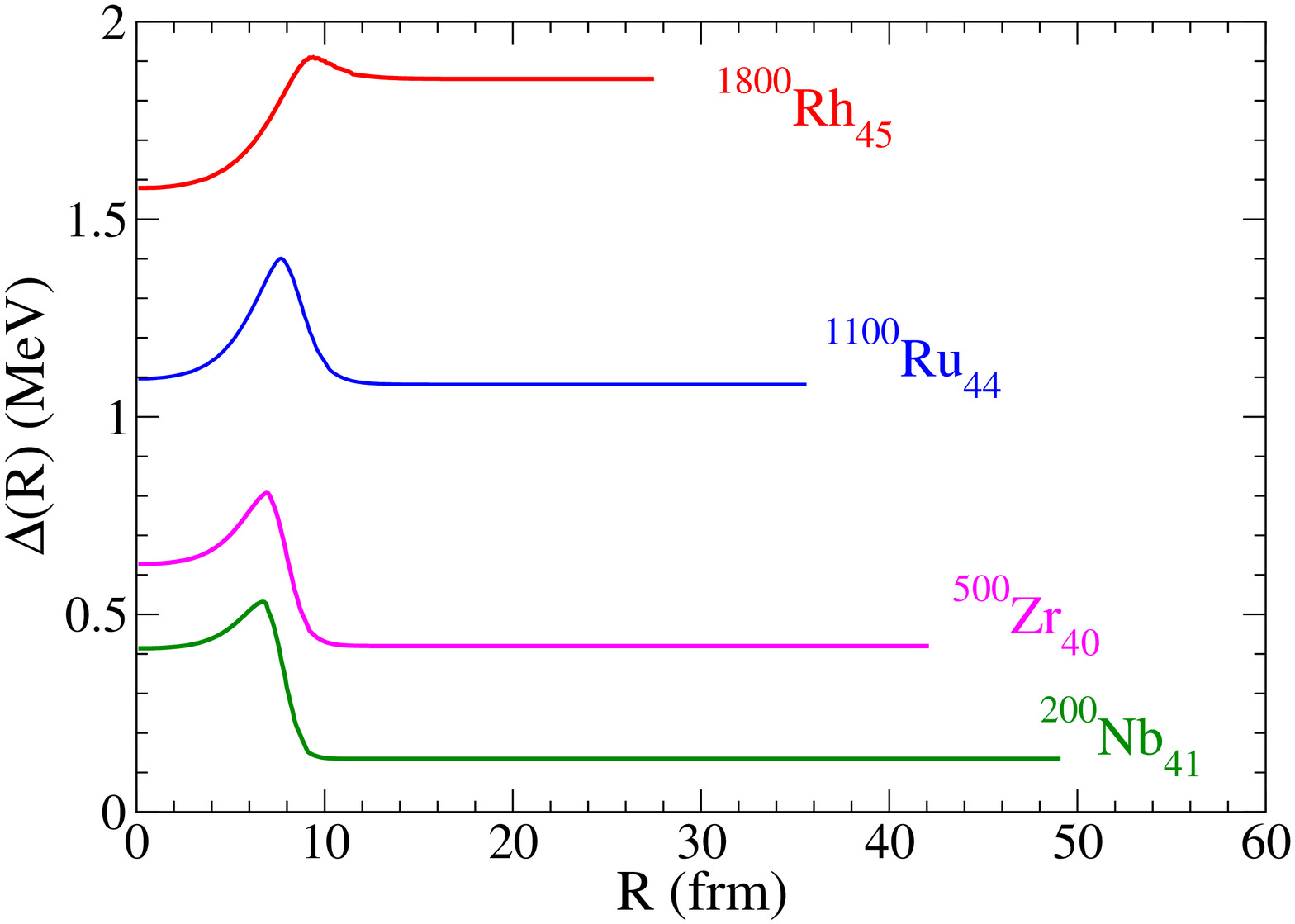}
\includegraphics[height=7.5cm,angle=-90]{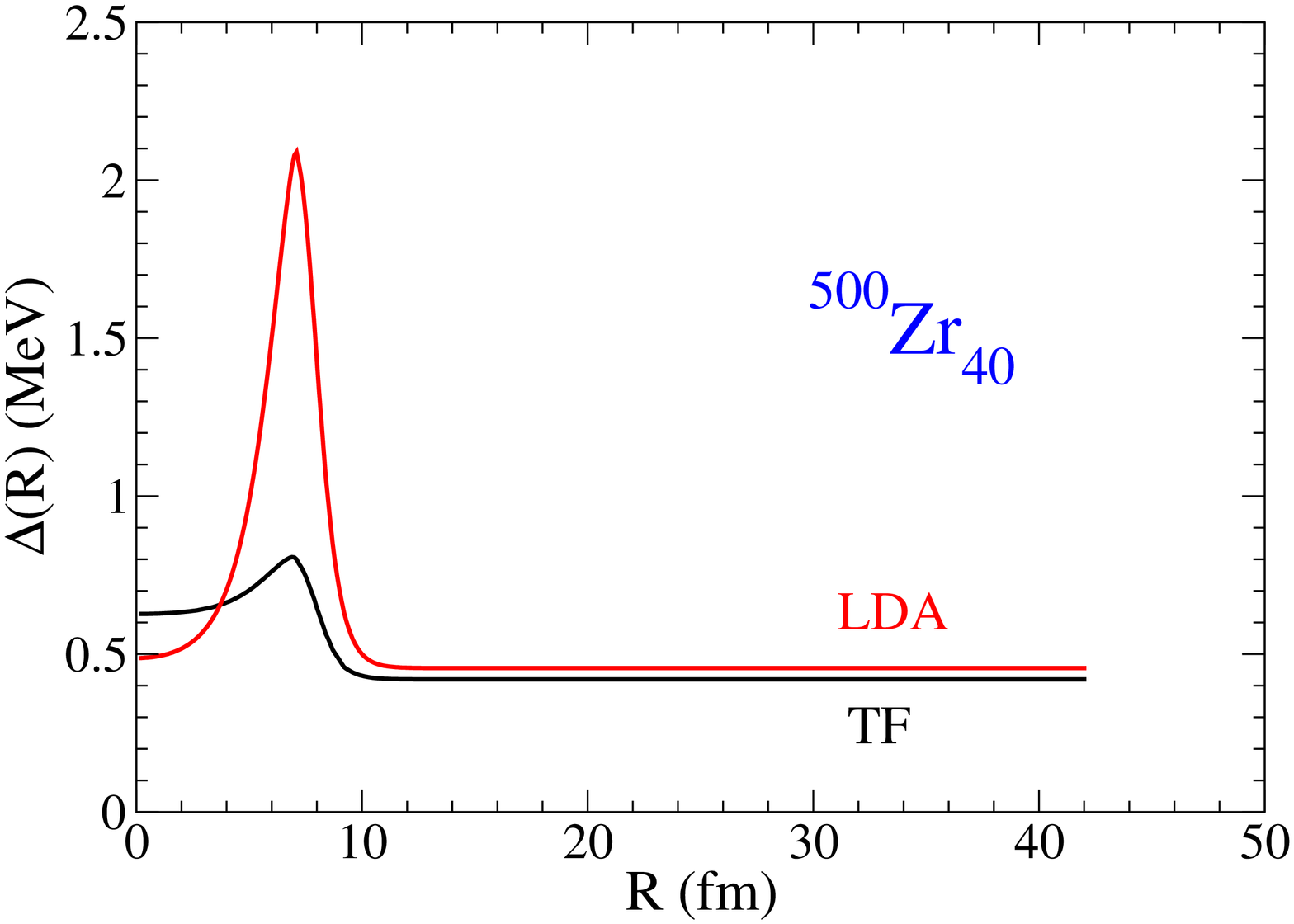}
\caption{(Coloronline) \label{Figure5} Upper panel: Radial dependence of the TF gap in the 
considered WS cells. The end points indicate the radius of the WS cells.
Lower panel: Comparison between TF and LDA gaps as a function of the 
position in a WS cell containing a single $^{500}_{40}$Zr nucleus.}
\end{figure}

We are now in a position to solve the quantal and TF gap equations for the 
above mentioned parameter values of our model. The result for the gap at the 
chemical potential $\mu$ is shown in Fig. 2 as a function of $\mu$. We start 
with $\mu$ from the bottom of the pocket well, i.e. with zero density. 
We then increase $\mu$, i.e. the density. We see that once the fill up of the 
pocket reaches its top, the values of the gap sharply drop and practically 
reach 
zero. In the continuum the gaps slowly rise again. We see that quantal and TF 
values are in close agreement. The overshoot of the TF solution for negative 
$\mu$ very likely is due to the smallness of the pocket which only can 
accomodate nine bound levels.  It may be partially cured including $\hbar$ 
corrections \cite{RS}
which, however, 
we do not consider here. Before we come to an explanation of the 
drop of the gaps at overflow (drip), let us study 
the gaps as a function of position in transverse direction: 
$\Delta(z) = -g K(z)$ with $K(z) = \sum K_n |\varphi_n(z)|^2$. 
Semiclassically, this expression becomes:
 $K(z) = \int_{V_0}^{\Lambda} dE g^{TF}(E) 
K(E) \rho^{TF}_E(z) $. 

In Fig.~3, we show the density profiles for three values of $\mu$: 
$\mu$ = 40, 0.5, and - 5 MeV. We see that quantal and TF results agree, 
up to shell fluctuations, very well. We also show the LDA results. We see that 
they can be as wrong as by 50 percent. For other choices of system parameters 
the LDA error may even be worse. This stems from the fact that in TF (and, 
of course, also quantally), there is coupling between inside and outside the 
pocket, i.e. the Cooper pair wave function extends into both regions what 
tends to equilibrate the values of the gaps. In LDA the contrast is much 
too strong. The drop of the gaps when crossing the threshold can be explained 
by the fact that the single particle states are strongly delocalised in the 
outer container and, thus, their contribution to the pairing matrix element 
$V_{n,n'}$ becomes very small.


Having gained faith into our TF approach, we now can explore other 
geometries and other systems, 
more difficult for quantal solutions. In Fig. 4 we show the result 
for $\Delta$ in the spherical double harmonic oscillator potential shown in 
Fig. 1b which may be realised with cold fermionic atoms to study the overflow 
situation. 
A zero range force with strength $g$=-1.0 and cut off $\Lambda$=164.34  
(in the corresponding optical trap units with $\omega_{opt}=2\pi \times 1000$ 
Hz taken from \cite{viv01}) is used. 
We see that the result is qualitatively similar to the slab 
case though in 
this spherical geometry the dip does not quite reach zero and also is shifted 
slightly to an energy above the break of the potential. Note that 
this 
depends strongly on the choice of the ratio $\omega_{mag}/\omega_{opt}$
as it can be seen in the figure. 
Also the gap starts to decrease 
towards the minimum quite early. It shall be interesting to see whether our 
prediction can be verified experimentally.

Let us now make a more realistic study of Wigner-Seitz (WS) cells
including electrons in $\beta$ equilibrium to simulate the inner crust of 
neutron stars  \cite{hae07}. 
To this end 
the mean-field is computed selfconsistently using the BCP energy density 
functional \cite{bal08} together with the TF approach as explained 
in \cite{vin11b}. 
The semiclassical description of the WS cells including pairing correlations 
at TF level is obtained from this mean-field using the finite range part 
of the Gogny D1S force \cite{D1S} in the pairing 
channel \cite{D1Sa}.
It must be pointed out that the total energy per baryon obtained with our
TF approach is in very good agreement with the old quantal calculation of 
Negele and Vautherin \cite{neg73} as it is explicitly discussed 
in Ref.\cite{vin11b}. 
In Fig. 5 are displayed the corresponding gaps with their radius dependences. 
It is seen that when the gap is small outside the region of the nucleus, 
then the gap 
also is small inside the nucleus. This stems from the very large 
coherence length where one neutron of a Cooper pair can be in the huge 
volume of the gas and the other inside the small volume of the 
nucleus (proximity effect). In this way the gas 
imprints its behavior for the gap also inside the nucleus. Such a conclusion 
was also given in a quantal Hartree-Fock-Bogoliubov (HFB) calculation 
by Grasso et al. in \cite{gra08} what shows
that the here employed BCS approximation  
apparently yields very similar answers as a full HFB calculation for WS 
cells \cite{bal07,HFB}. Finally, in the lower panel of Fig. 5 we show a 
comparison of LDA and 
present TF results for a particular WS cell. 
We see a huge difference in the surface region of the nucleus. This simply 
stems from the fact that in this case of the $^{500}_{40}$Zr nucleus in the WS 
cell the gap is very small and, therefore, the coherence length very large 
invalidating LDA. A study with examples a little less unfavorable for LDA is 
given in \cite{pas08}.

For isolated nuclei at the neutron drip the situation may be different. 
It seems that 
in this situation the difference between HFB and BCS approaches may  
be significant. Somewhat conflicting results in this respect exist in the 
literature. In ref \cite{ham05} very similar results to ours are found 
for S-wave pairing. On the other hand in \cite{taj05} 
the gap seems to rise towards the drip before it bends down. Similar results 
have recently been found in \cite{hag11}. Preliminary investigations 
show that these 
discrepancies may be due to large shell fluctuations in isolated nuclei. 
More studies in this direction seem to be necessary.


Summarizing, we have studied superfluid fermions in a large container, either 
external (cold atoms) or created self consistently (nuclei) for situations 
where the top of the fluid reaches the edge of a small pocket situated at the 
origin of the wide confining potential. The gap drops to zero at the edge 
before rising again when 
the density fills up the outer container. This at first somewhat surprising 
phenomenon can be explained quite straightforwardly. Such situations, as 
already mentioned, can exist in cold atoms and
nuclei in the inner crust of neutron stars, two examples treated here with 
their specific form of containers. For small systems like isolated nuclei at 
the neutron drip, the situation may be blurred by shell effects.

As an important second aspect of this work, we showed that a novel 
Thomas-Fermi approach to 
inhomogeneous situations can cope with situations where LDA 
fails. This means that our TF approach is free of the restrictive condition, 
prevailing for LDA, that the Cooper pair coherence length must be shorter 
than a typical length $l$  
(the oscillator length in the case of a harmonic container) over 
which the mean field varies appreciably. On the contrary, 
our TF theory has the usual TF validity criterion, namely that local 
wavelengths must be shorter than $l$. 

The accuracy 
of our TF approach opens wide perspectives for a treatment of 
inhomogeneous superfluid Fermi-systems with a great number of particles 
not accessible for a quantal solution of the BCS (HFB) equations. Such 
systems may be cold atoms in deformed containers (eventually reaching millions 
of particles), superfluid-normal fluid (SN) interfaces, vortex profiles, etc. 
As a matter of fact, as is well known \cite{RS}, the TF approach becomes the 
more accurate, the larger the system. Thus the TF approximation is 
complementary 
to the quantal one in the sense that the former works where the latter is 
difficult or even impossible to be obtained numerically.

We thank W. Ketterle and A.  Minguzzi for their interest in this work and 
for pointing out refs. 
\cite{kett98,viv01}. We are greatful to Michel Farine for contributions and to 
Michael 
Urban for useful 
discussions and a critical reading of the manuscript.
We also thank K. Hagino for pointing to Ref.\cite{taj05} and sending their own 
results prior to publication.
This work has been partially supported by the IN2P3-CAICYT 
collaboration (ACI-10-000592). 
One of us (X.V.) acknowledges grants FIS2008-01661 (Spain and FEDER),
2009SGR-1289 (Spain) and Consolider Ingenio Programme CSD2007-00042 for 
financial support.


\begin{thebibliography}{99}
\bibitem{D1Sa}
The D1S force for pairing has been multiplied with
a factor 0.85 to compensate for the use of the bare mass in the calculation.
\bibitem{HFB}
Our TF theory can be generalised to the HFB case what shall be done
in future work.
\end{thebibliography}
\end{document}